\documentclass[aps, pra, twocolumn, amsmath, graphicx, latexsym, asmsymb, amsfonts]{revtex4}

\usepackage{epsfig}
\usepackage{hep}
\usepackage{graphicx}
\usepackage{moreverb}
\usepackage{fancyhdr}
\usepackage{supertabular}
\usepackage{color}

\usepackage{amssymb}

\begin{document}

\title{Error-Resistant Distributed Quantum Computation in Trapped Ion Chain}

\author{Sibylle Braungardt$^{1}$, Aditi Sen(De)$^{1}$,
Ujjwal Sen$^{1}$,
and Maciej Lewenstein$^{2}$}

\affiliation{\(^1\)ICFO-Institut de Ci\`encies Fot\`oniques,
Mediterranean Technology Park, 08860  Castelldefels (Barcelona), Spain\\
\(^2\)ICREA and ICFO-Institut de Ci\`encies Fot\`oniques,
Mediterranean Technology Park, 08860 Castelldefels (Barcelona),
Spain}

\begin{abstract}

We consider experimentally feasible chains of trapped ions with
pseudo-spin 1/2, and find models that can potentially be used to
implement error-resistant quantum computation. Similar in spirit
to classical neural networks, the error-resistance of the system
is achieved by encoding the qubits \emph{distributed} over the
whole system. We therefore call our system a \emph{quantum neural
network}, and present a \emph{quantum neural network model of quantum computation}.
Qubits are encoded in a few quasi-degenerated low
energy levels of the whole system, separated by a large gap from
the excited states, and large energy barriers between themselves.
 We investigate
protocols for implementing a universal set of quantum logic gates
in the system, by adiabatic passage of a few low-lying energy
levels of the whole system. Naturally appearing and potentially dangerous distributed
noise in the
system leaves the fidelity of the computation virtually unchanged,
if it is not too strong. The computation is also naturally resilient to local perturbations
of the spins.

\end{abstract}
\maketitle

\section{introduction}

Quantum computers, if realized in laboratory, are known to be
capable of solving problems much faster than classical computers.
Two famous examples are the Shor algorithm \cite{shor} for factoring
a nonprime integer \(N\) in polynomial time in the number of binary
digits of \(N\), and the Grover algorithm \cite{grover}, which can
find a single object from an unsorted database of \(N\) objects in
an order of
$\sqrt{N}$ calls to the database in a quantum computer.  While
the latter task requires an order of \(N\) calls to the database in a
classical computer, the former is strongly believed to require
exponentially large time in the same.

One of the most
challenging problems that occur when trying to build a quantum
computer is decoherence. The system interacts with its environment,
and the quantum logical gates cannot  be implemented perfectly. A number of schemes
for protecting quantum information have been developed, including
fault tolerance codes \cite{faulttol}, decoherence free subspaces
\cite{decfreesubsp}, noiseless subsystems \cite{noiselessubsys},
dynamic decoupling \cite{dyndecoupling}, topological quantum
computation \cite{topological}, and geometric quantum computation \cite{gyamiti}.

Our approach to error resistant quantum computation is based on
the idea of neural networks, which, classically, can offer robust
(i.e. noise resistant) storage and manipulation of classical data
by encoding the classical memory patterns in a \emph{distributed}
way in the
whole neural network (see e.g.
\cite{neuralnets}). A typical classical neural network has a large
number of metastable energy minima with large basins of attraction,
which can be used for this purpose. A classical neural network is
also typically characterized by long range interactions.
Moreover,
these interactions are usually disordered and ``frustrated''. The
\emph{disordered} interactions are motivated by realistic
situations: The bonds that carry information between neurons in a
brain are typically quite irregular, and fluctuate.
Such disordered interactions have the effect that the different metastable
energy minima are statistically independent, so that for large systems, their overlaps vanish.
``Frustration'' in a network can
be defined as a situation, where one cannot find a configuration
of the ``particles'' (that make up the network) by satisfying all
the interactions (bonds) between them.  While there are physical
(or biological) reasons for considering frustrated interactions,
it is also (believed to be) important for having a large number of
low lying metastable
and ``orthogonal'' (in the sense of Hamming distance (see e.g. \cite{eita-Hamming})) energy patterns.

Just as distributed classical information encoding in classical
neural networks is good for classical data manipulation, we show
that \emph{distributed quantum information encoding} in their
quantum analogs (we call them ``quantum neural networks'' (QNN))
can potentially be used for robust manipulation of quantum data:
error resistant quantum computation.
 The system that we have in our
minds for a possible implementation of the  protocols that we
describe in this paper, are systems of cold ions in a trap (see
\cite{Porras-Cirac, Wunderlich, review} and references therein).
The state-of-the-art of current experiments (see e.g.
\cite{eita-Blatt, eita-Wineland_at_Boulder}, and references
therein) show, that such systems allow for
a high degree of control of the parameters, and in
particular, of the interactions. Consequently,
in such systems, we are able to manipulate strictly orthogonal (in
the usual sense of orthogonality of pure states in a Hilbert
space) states of the whole system, without making use
of disordered interactions.
Moreover,  this is possible
with
a
mesoscopic
number of ions in the system.

We propose to encode
quantum data in the energy levels of the system, and perform
quantum gates by adiabatic passage of these levels. Thus, a too large
number of low lying energy levels will typically be detrimental
for our purposes: the finer an avoided crossing is, the larger is
the probability of the system to leak into higher excitations.
Therefore, we also do not want  frustration effects to dominate
in our system and produce such low energy states.

Using such a quantum neural network, we show that one can
implement not only one-qubit gates, but also universal two-qubit
gates in a naturally error resistant way. The idea of the gate
implementations is the following. Suppose that a (unitary) gate is
defined as a transfer of an initial orthogonal set of vectors into
a final one. We choose the initial parameters of the Hamiltonian
of the system in such a way, that the initial orthogonal set of
vectors can be encoded onto a few lower eigenstates of the initial
Hamiltonian. Subsequently, the system parameters are changed
(slowly, i.e. adiabatically), such that the final orthogonal set
of vectors of the unitary gate, turns out be (approximately) the
corresponding lower instantaneous eigenstates of the final
Hamiltonian. The change in the Hamiltonian is brought about by
the changing of certain external (parallel and transverse) fields, and
these are the sole (external) parameters that needs to be changed
for the adiabatic passage.

In a certain sense, our system resembles systems with topological
order and topologically protected qubits, such as proposed in Ref.
\cite{topological}. In those systems there exist several
degenerated ground states that are separated from the excited
states by a large energy gap and which are used to store the
qubits. Local perturbations of such states are very insufficient:
one has to go to very high orders of the perturbation theory in
order to transfer one of the protected states into another, i.e.
change the topological charge.

In our case, the situation is rather similar for most practical purposes.
As we showed in Ref. \cite{ager-paper-abar}, we can
control the ion trap potential in such a way that the system has
 a quasi-degenerate low energy manifold. As the classical analysis in our earlier
Letter (Ref. \cite{ager-paper-abar}) shows, all of these states are local minima of energy, which are
stable with respect to multiple spin flips, just as in standard classical
neural networks. We therefore have a number of states that are
separated by a large gap from the higher excited states, and are
separated between themselves by large barriers. This implies that also in our case,
local perturbations  are inefficient. The system
(QNN) is thus \emph{intrinsically} robust to local noise for
quantum computational purposes - qubit states, when slightly perturbed
locally, transform under gate operations, almost similarly as ideal qubit
states. On top of that, the system exhibits even a different
mechanism of error resistance, as we shall see below. In constructing
the QNN, in a natural manner one obtains noise terms, which have a
distributed character: they act globally on the whole system. Such
terms are potentially dangerous and may seriously diminish the
performance of the model. Fortunately, also in this case, energy
gaps and barriers assure protection.

The paper is organized as follows. In Sect. \ref{sec-adtheorem}, we
briefly describe the adiabatic theorem. In Sect. \ref{sec-QNN
hamiltonian},  we give a description of the model of our QNN and
introduce our noise models: local perturbations and
the naturally appearing distributed noise.
The \emph{quantum neural network model of quantum computation} consists of two main steps: The distributed encoding of the qubits, and
the implementation of the quantum gates by adiabatic passage.
The encoding of the
qubits is described in Sect. \ref{sec-qubit}. Sect. \ref{sec-gates}
defines the two gates, namely the \({\cal H}\) gate and the Bell
gate, whose protocols for implementation are presented in Sect.
\ref{sec-protocols}. Sect. \ref{sec-fidelity} contains the
resulting fidelities of the gates. In Sect. \ref{sec-adiabaticity},
we apply the adiabaticity condition to our system, and give
constraints on the time of the evolution. We discuss our results in Sect. \ref{sec-summary}.

\section{The adiabatic theorem}
\label{sec-adtheorem}

The quantum adiabatic theorem \cite{born,messiah} states that a
physical system that is initially in one of its nondegenerate
eigenstates
 will remain in the corresponding instantaneous eigenstate,
provided that the Hamiltonian is varied ``sufficiently'' slowly.

The time evolution of the system is given by the time dependent
Schr{\" o}dinger equation
\begin{equation}
\label{schrodinger} i\hbar\frac{d}{dt}\ket{\Psi(t)} = H(t)\ket{\Psi(t)},
\end{equation}
where we let our system evolve from $t=0$ until the time
$t=T$.
If we scale the
time evolution by introducing a scale factor $s=\frac{t}{T}$, where
$0\leq s\leq 1$, the Schr{\" o}dinger equation
becomes
\begin{equation}
\label{ei_eq_bujhtey_hobey} i\hbar\frac{d}{ds}\ket{\Psi(s)} = T
H(s)\Psi(s).
\end{equation}
The time evolution of the system is
described
completely
by
the
Hamiltonian
and the initial state. The development of the system is considered as
``adiabatic'', so that the adiabatic theorem holds, if the change of the Hamiltonian is small as compared to
the gap \(g(s)\) between the energy levels; more precisely, if
\begin{equation}
\label{eq-roti-kapra-aur-makaan}
T \gg \hbar
\frac{\parallel \frac{d}{ds}H(s) \parallel}{g(s)^2},
\end{equation}
where \(\parallel \Lambda \parallel\) is the operator norm of
\(\Lambda\), defined as the square root of the maximal eigenvalue
of  \(\Lambda^\dagger \Lambda\). If one desires to  adiabatically
transport the \(i\)th eigenstates at a certain time to the \(i\)th
eigenstate at a different time,  the gap \(g(s)\) is the minimum
of the energy gaps to the \((i-1)\)th and the \((i+1)\)th energy
levels. If the adiabaticity condition is fulfilled, an evolution
starting out in the \(i\)th eigenstate  of H(0) will end up, at
time \(t=T\), with high probability, in the \(i\)th eigentstate of
the Hamiltonian $H(T)$.

In this paper, adiabatic transport of superpositions of a few
energy levels is considered. For a superposition of say the 2nd
and the 3rd levels, the gap \(g(s)\) is the minimum of the gaps
between 1st and 2nd levels, 2nd and 3rd levels, and 3rd and 4th
levels. When applying the adiabatic theorem to superpositions of
eigenstates, the phases are also relevant to the calculations. For
example, a superposition
\begin{equation}
a|2(0)\rangle + b |3(0)\rangle
\end{equation}
of the
2nd energy level \(|2(0)\rangle\) and the 3rd level
\(|3(0)\rangle\) of the Hamiltonian \(H(0)\), will end up, at time
\(T\), with high probability, in the superposition
\begin{equation}
ae^{i\Phi_2}|2(T)\rangle + be^{i\Phi_3} |3(T)\rangle
\end{equation}
of the
2nd energy level \(|2(T)\rangle\) and the 3rd level
\(|3(T)\rangle\) of the Hamiltonian \(H(T)\). The phases $\Phi _i$
are given by the sums of the dynamical and Berry phases
\cite{Pancharatnam-birat-byapar, Berry, Wilczek, messiah} for the
corresponding eigenstates. The dynamical phase is given by
\begin{equation}
\Phi_i^D = -\int_{0}^{T}E_i(t)dt,
\end{equation}
and the Berry phase is defined as
\begin{equation}
\Phi_i^B = i\int_{0}^{T}\bra{E_i(t)}\frac{d}{dt}\ket{E_i(t)}dt.
\end{equation}
The instantaneous eigenvalues of the Hamiltonian $H(t)$ are
denoted as $E_i(t)\)  \((i=0,1,2,\ldots)$, with $E_0(t)<E_1(t)<E_2(t)<\ldots$, for all time \(t\).
The instantaneous $i$th eigenstate is denoted as $\ket{E_i(t)}$.
The ground state $\ket{E_0(t)}$ will also be denoted as
$\ket{G(t)}$.

Since the work of Farhi and Gutmann \cite{farhi} (see also
\cite{aro_onekei_achhey, ina-mina-dika}, and references therein),
the adiabatic theorem has been used for quantum information
processing, and has been called ``adiabatic quantum computation''.
A methodological difference between the above set of works and the
present paper, is that in their case, the system is always in the
ground state, while our system is typically a superposition of a
few lower excited levels along with the ground state. Among other
things, this may affect the adiabaticity condition. Perhaps even
more important differences are as follows:
\begin{itemize}
\item[(i)] \emph{``Special purpose'' Hamiltonian versus ``universal'' Hamiltonian}: Adiabatic quantum computation typically considers
 a certain quantum algorithm, and depending on the algorithm, a certain Hamiltonian is considered. It was shown
in Ref. \cite{ina-mina-dika} that the set of 2-local Hamiltonians is enough for this purpose.
We, however, have a single quantum Hamiltonian (the QNN), that we will show below to be enough  for all quantum algorithms, as
our Hamiltonian implements universal gates (like the Bell gate, defined in Sect. \ref{sec-gates}),
which can be applied to simulate
arbitrary
quantum algorithms. In this sense, the QNN Hamiltonian is a \emph{universal} Hamiltonian for quantum computation.

\item[(ii)] \emph{Noise-resistance mechanism}:
Below, we will observe that quantum computing in a system described by the QNN Hamiltonian is resistant to noise,
and this resistance is related to the fact that the system mimics a neural network: the quantum information is distributed
in the eigenstates of the whole system. Resistance to noise in adiabatic quantum computation has apparently a different origin, as
the typical Hamiltonians there, are not fully connected \cite{ina-mina-dika}.

\end{itemize}

\section{The quantum neural network hamiltonian and our noise models} \label{sec-QNN hamiltonian}

In this paper we will consider a system of trapped spin-1/2
particles with long range interactions, that are subject to slowly
changing (in real time (\(t\)))  external magnetic fields. Such a
system can be implemented with ions in a trap, where two internal
states of each ion serve as the ``up'' and ``down'' states (denoted as \(\ket{\uparrow}\) and \(\ket{\downarrow}\) respectively) of the
pseudo spin-1/2 particles (see Refs. \cite{Porras-Cirac,Wunderlich}). As
shown in the above references, such a system offers a wide variety
of spin models, which can be implemented by changing the system
parameters. We are interested in long range Ising interactions. As
shown in Refs. \cite{ager-paper,ager-paper-abar}, the Hamiltonian
of the system depends crucially on the geometry of the external
trap potential. For the case of a harmonic trap, the time
dependent Hamiltonian of a system of eight spins can be
approximated by
\begin{align}
H(t) = - \lambda \Big[ r_1 \left(S_{z1}+S_{z2}+S_{z3}+S_{z4}\right)^2 \nonumber \\
+r_2\left((S_{z1}+S_{z2})-(S_{z3}+S_{z4})\right)^2 \nonumber \\
+ r_3((S_{z1}-S_{z2})-(S_{z3}+S_{z4}))^2 \nonumber \\
+A(t)(S_{x1}+S_{x2}+S_{x3}+S_{x4}) \nonumber \\
+B_1(t)(S_{z1}+S_{z2})+B_2(t)(S_{z3}+S_{z4})\Big], \label{eqhamilton}
\end{align}
where, typically, \(r_1\) is much greater than \(r_2\) and
\(r_3\). The \(r_i\) corresponding to higher modes are even
smaller and are thus neglected.
Here
\begin{equation}
S_{\alpha i}=\sigma^{\alpha}_{2i -1} + \sigma^{\alpha}_{2i},
\mbox{ }i=1,2,3,4,
\end{equation}
 and $A$, $B_1$ and $B_2$ are external
magnetic fields.
  The overall factor \(\lambda\), which has the
units of energy, in the Hamiltonian
\(H(t)\) has the effect of making the rest of the parameters in the Hamiltonian  dimensionless.
As we will show, such a system (i.e., one in which \(r_1 \gg r_2, r_3\)) can be used for
implementing one-qubit gates, but is apparently not suitable for two-qubit universal gates. However, for trap potentials
 of the form \(|x|^\gamma\),
with \(\gamma \approx 0.5\), one obtains a situation where \(r_1
\approx r_2 \gg r_3\) \cite{ager-paper,ager-paper-abar}. The trap for which \(\gamma =0.5\) may be called a \emph{fountain trap}.
We show
below that this latter case can be used for implementing both one-
qubit and two-qubit gates. The consideration of eight spins in our
system is motivated by the number of spins that is currently
viable in ion trap experiments (see e.g. \cite{eita-Blatt}).

We will now discuss two possible sources of noise that act on the
system. We first consider a ``distributed noise'', i.e. noise that
arises globally in the system. Secondly, we will discuss a ``local
noise'', i.e. noise that acts on the states of the system via local perturbations.

\subsection{Distibuted Noise}

The terms in the quantum neural network Hamiltonian \(H(t)\)
corresponding to \(r_1\), \(r_2\), and \(r_3\) (\(r_4\), \(r_5\),
etc. are neglected here) stem respectively from the first, second,
and third (and further) vibrational modes of the trapped ions
system, since the phonons are the carriers of interactions between
the spins. Therefore, in the case when \(r_1 \approx r_2 \gg
r_3\), one can consider the \(r_3\) (as well as \(r_4\), \(r_5\),
etc., if present) term as a distributed, and thus potentially
dangerous,   noise in the system. This noise model is motivated by
taking into account the following points:

\begin{itemize}
\item[(i)] Increasing the effect of the third (and also higher) vibrational
mode, which in the undisturbed case is  much smaller than the
first and second ones, covers inaccuracies in the trapping
potential.

\item[(ii)] Moreover,  decreasing the eigenfrequency of the
third vibrational mode (i.e. increasing \(r_3\)) introduces a disturbance in the motion of
the ions.

\item[(iii)] In addition, this introduces noise in the spin, as it is the
phonon modes that are the carriers of interaction between the
effective spins.

\end{itemize}

Similarly, in the case when $r_1 \gg r_2, r_3$ (e.g. in the case of the harmonic trap), the $r_2$ term can
be considered as a model of noise in the system.

\subsection{Local Noise}
We will now discuss our model for local noise that can potentially act on our system, by considering local
perturbations to the states of the system. 
Let us note that the term ``local noise'' in this manuscript does not mean that the noise is local 
with respect to the spins (ions). It means that the noise is \emph{local in the configuration space} of the system. 
We consider
imperfections to the initial state of the system by superposing it
with the states where one spin is flipped. For the ``all down''
state \(\ket{\downarrow\downarrow\downarrow\downarrow\downarrow\downarrow\downarrow\downarrow}\), for example, we
consider the transformation
\begin{equation}
\ket{\downarrow\downarrow\downarrow\downarrow\downarrow\downarrow\downarrow\downarrow}
\rightarrow
\ket{\downarrow\downarrow\downarrow\downarrow\downarrow\downarrow\downarrow\downarrow}+\varepsilon
\ket{W_8},
\end{equation}
as a model of noise. The output of the noise effect is not yet normalized.
Here $\ket{W_8}$ is the eight spin W-state, defined as the normalized symmetric
superposition of all states with one spin up and the rest spins down \cite{W-state}. The parameter
 $\varepsilon$ is the strength of the perturbation. In Sect.
\ref{sec-fidelity}, we will study the influence of this local noise on
the gate fidelities.

\section{Distributed encoding of the qubits} \label{sec-qubit}

As noted before, the quantum neural network model of quantum computation consists of two steps, beginning with
an encoding of the qubits in a distributed way: The qubits are encoded, as we discuss now, as eigenstates of the whole neural network.

We assume that  the Hamiltonian \(H(t)\) changes in a continuous
way from a certain initial value $H(0)$ at time $t=0$ to a certain
final value $H(T)$ at time $t=T$. Note that the change in the
Hamiltonian is brought about solely by changes in the fields. We
choose the initial fields in the QNN Hamiltonian such that the
ground state and the three lowest excited states at the initial
time \(t=0\) are respectively
\begin{eqnarray}
\ket{G(0)} &=& \ket{\uparrow \uparrow \uparrow \uparrow \uparrow \uparrow \uparrow \uparrow }, \nonumber \\
\ket{E_1(0)} &=& \ket{\downarrow \downarrow \downarrow \downarrow \downarrow \downarrow \downarrow \downarrow }, \nonumber \\
\ket{E_2(0)} &=& \ket{\uparrow \uparrow \uparrow \uparrow \downarrow \downarrow \downarrow \downarrow }, \nonumber \\
\ket{E_3(0)} &=& \ket{\downarrow \downarrow \downarrow \downarrow \uparrow \uparrow \uparrow \uparrow }.
\end{eqnarray}

For implementing one-qubit gates, we will use the following encoding:
\begin{eqnarray}
\label{eq-ek}
\ket{0} = \ket{G(0)} &=& \ket{\uparrow \uparrow \uparrow \uparrow \uparrow \uparrow \uparrow \uparrow }, \nonumber \\
\ket{1} = \ket{E_1(0)} &=& \ket{\downarrow \downarrow \downarrow \downarrow \downarrow \downarrow \downarrow \downarrow }.
\end{eqnarray}
The extreme left hand sides of the above equations denote the logical states of the qubit.

On the other hand, for two-qubit gates, we will encode one qubit
in the first four spins and the other qubit in the remaining four
spins:
\begin{eqnarray}
\label{eq-dui}
\ket{00} = \ket{G(0)} &=& \ket{\uparrow \uparrow \uparrow \uparrow \uparrow \uparrow \uparrow \uparrow }, \nonumber \\
\ket{11} = \ket{E_1(0)} &=& \ket{\downarrow \downarrow \downarrow \downarrow \downarrow \downarrow \downarrow \downarrow }, \nonumber \\
\ket{01} = \ket{E_2(0)} &=& \ket{\uparrow \uparrow \uparrow \uparrow \downarrow \downarrow \downarrow \downarrow }, \nonumber \\
\ket{10} = \ket{E_3(0)} &=& \ket{\downarrow \downarrow \downarrow \downarrow \uparrow \uparrow \uparrow \uparrow }.
\end{eqnarray}
Again the extreme left hand sides of the above equations denote the logical states of the two qubits.

\section{The \({\cal H}\) gate and the Bell gate}
\label{sec-gates}

We consider implementations of a one-qubit, as well as a two-qubit
gate. The two-qubit gate is an entangling one, so that along with
one-qubit gates, a universal set of quantum gates is formed
\cite{phuler-upottyoka}.
The one-qubit gate, in the logical basis, is given by
\begin{equation}
\ket{0} \rightarrow \ket{+} \equiv \frac{\ket{0}+\ket{1}}{\sqrt{2}}, \,\,\,\,\,\,\, \ket{1} \rightarrow -\ket{-} \equiv
-\frac{\ket{0}-\ket{1}}{\sqrt{2}}.
\end{equation}

Note that this transformation, which we call the \({\cal H}\) gate, is closely related to
the Hadamard transformation that takes
\begin{equation}\ket{0} \rightarrow
\ket{+} \mbox{ and } \ket{1} \rightarrow \ket{-}.
\end{equation}

The two-qubit gate that we consider here
acts as
\begin{equation}
\ket{00}\rightarrow \ket{\phi^+} \equiv \frac{\ket{00}+\ket{11}}{\sqrt{2}},
\end{equation}
\begin{equation}\ket{11}\rightarrow -\ket{\phi^-} \equiv
-\frac{\ket{00}-\ket{11}}{\sqrt{2}},
\end{equation}
\begin{equation}\ket{01}\rightarrow \ket{\psi^+} \equiv \frac{\ket{01}+\ket{10}}{\sqrt{2}},\end{equation}
\begin{equation}\ket{10}\rightarrow - \ket{\psi^-} \equiv -\frac{\ket{01}-\ket{10}}{\sqrt{2}}.\end{equation}
The gate is manifestly entangling, and we call it the Bell gate.

\section{The gate implementation protocols: Adiabatic Passage} \label{sec-protocols}

The second and final step of our quantum neural network model of quantum computation consists in the implementation of the quantum gates,
by adiabatic passage of the whole system by adiabatic tuning of parallel and transverse magnetic fields.

\subsection{Protocol for the \({\cal H}\) gate}

\begin{figure}[t]
\begin{center}
\includegraphics[width=0.9\linewidth]{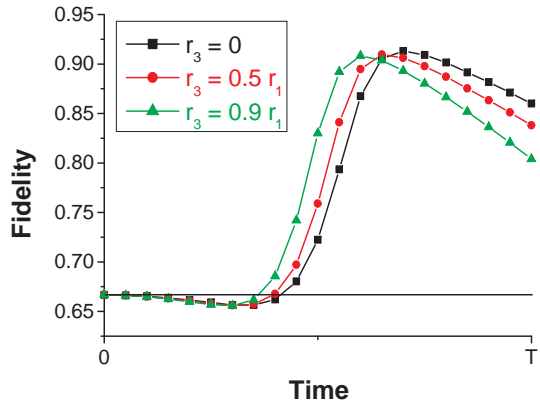}
\caption{(Color online.) Fidelity of the \({\cal H}\) gate, as a function of time: Effect of distributed noise.
The fidelities are calculated for
\(r_1 = 10\), and \(r_2 = 9.5\), for different values of \(r_3\). The (parallel and transverse) fields
for which the calculations are performed are depicted in Fig. \ref{fig-fields}. As seen in the figure, the maximal fidelities are obtained
a little after \(t=T/2\). \(T\) is a time that satisfies Eq.
(\ref{eq-roti-kapra-aur-makaan}), which with our chosen parameters mean
\(T \gg 7 \times 10^6 \hbar/\lambda\).
 Note that the fidelity does not change appreciably with the increase of the distributed noise level \(r_3\).
 The local noise is assumed to absent (i.e. \(\varepsilon =0\)).
The horizontal line at \(2/3\) denotes the limit above which the gate fidelity is quantum.
}
\label{fig-hadamard}
 \end{center}
\end{figure}

\begin{figure}[t]
\begin{center}
\includegraphics[width=0.9\linewidth]{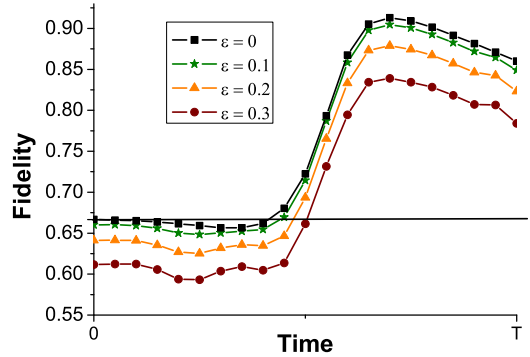}
\caption{(Color online.) Fidelity of the \({\cal H}\) gate, as a
function of time: Effect of local noise. The fidelities are
calculated for \(r_1 = 10\), \(r_2 = 9.5\), and \(r_3 = 0\), for
different values of \(\varepsilon\). Just like in the case of
distributed noise in Fig. \ref{fig-hadamard}, the (parallel and
transverse) fields for which the calculations are performed are
depicted in Fig. \ref{fig-fields}. And again, the maximal
fidelities are obtained a little after \(t=T/2\), where \(T\) is a
time that satisfies Eq. (\ref{eq-roti-kapra-aur-makaan}), which
with our chosen parameters mean \(T \gg 7 \times 10^6
\hbar/\lambda\). The fidelity does not show a marked diminish with
the increase of the level of local noise \(\varepsilon\). The
distributed noise is assumed to be absent (i.e. \(r_3=0\)). The
horizontal line at \(2/3\) denotes the limit above which the gate
fidelity is quantum. Note that the \(\varepsilon =0\) curve in
this figure is the same as the \(r_3=0\) curve in Fig.
\ref{fig-hadamard}. } \label{fig-local_noise}
 \end{center}
\end{figure}

\begin{figure}[t]
\begin{center}
\includegraphics[width=0.9\linewidth]{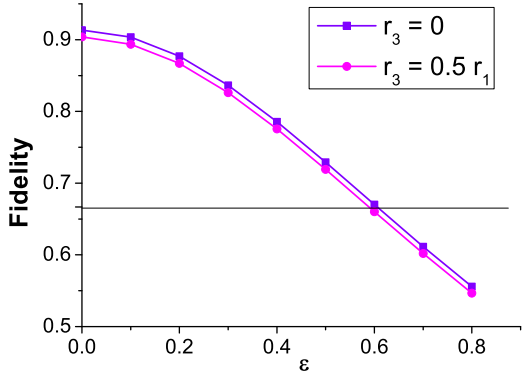}
\caption{(Color online.) Fidelity of the \({\cal H}\) gate:
Combined effect of distributed and local noise. We plot the
maximal fidelity of the \({\cal H}\) gate as a function of the
local noise parameter \(\varepsilon\), when the distributed noise
parameter \(r_3\) is vanishing and \(0.5\). The time for
calculation of the fidelities is chosen to be the one for which
maximal fidelity is obtained for \(\varepsilon =0\) and \(r_3=0\).
Again the fields are changed as in Fig. \ref{fig-fields}, and
adiabaticity requires that the time of fidelity calculation be
\(\gg 7 \times 10^6 \hbar/\lambda\). The horizontal line at
\(2/3\) denotes the limit above which the gate fidelity is
quantum. } \label{fig-loc+glob}
 \end{center}
\end{figure}

Let us first consider the protocol for implementing the single
qubit \({\cal H}\) gate. Note that in this case, the encoding is
given by Eq. (\ref{eq-ek}). To implement the \({\cal H}\) gate, a
qubit that is initially in the state
\begin{equation}a_0\ket{0}+a_1\ket{1}
\end{equation}
 (in
the logical basis), should evolve into the state
\begin{equation}
a_0\ket{+} -
a_1\ket{-}.
\end{equation}
 Here \(a_0\) and \(a_1\) are complex numbers, with
$|a_0|^2+|a_1|^2=1$. Using the encoding in Eq. (\ref{eq-ek}), the
qubit is initially in the state
\begin{equation}
\label{eq-haora}
a_0 \ket{G(0)} + a_1 \ket{E_1(0)}.
\end{equation}
We now adiabatically change the fields in the QNN Hamiltonian up
to a certain time \(t=T\), in which case, the system that was
initially in the state in Eq. (\ref{eq-haora}), evolves, in
accordance with the adiabatic theorem, to the state
\begin{equation}
\label{eq-champa} a_0e^{i\Phi _0}\ket{G(T)}+a_1e^{i\Phi
_1}\ket{E_1(T)},
\end{equation}
where the phases $\Phi _i$ are given by the sums of the dynamical
and Berry phases for the corresponding eigenstates
\cite{Pancharatnam-birat-byapar, Berry, Wilczek, messiah}.
The eigenvectors of the Hamiltonian that appear in our
calculations of the fidelities of the \({\cal H}\) gate as well as
the Bell gate, are all real in at least one basis. Consequently,
the corresponding Berry phases vanish. Therefore, the total phase
is given by the dynamical phase:
\begin{equation}\Phi_i=\Phi_i^D = -\int_{0}^{T}E_i(t)dt, \quad i=0,1,2, \ldots.
\end{equation}

Our aim is to change the fields in such a way that the final (time
evolved) state in Eq. (\ref{eq-champa}) is ``as close as possible''
 to the \({\cal H}\) rotated state
$a_0\ket{+}-a_1\ket{-}$, i.e. to
\begin{equation}a_0 \frac{\ket{G(0)} + \ket{E_1(0)}}{\sqrt{2}} - a_1 \frac{\ket{G(0)} - \ket{E_1(0)}}{\sqrt{2}}.
\end{equation}
The measure of closeness that we use is described in Subsect. \ref{subsec-nikot}).

\begin{figure}[t]
\begin{center}
\includegraphics[width=0.9\linewidth]{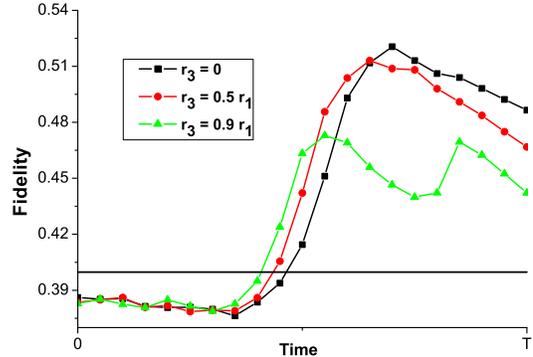}
\caption{(Color online.) Fidelity of the Bell gate, as a function of time. Just like in the case of the
\({\cal H}\) gate in Fig. \ref{fig-hadamard}, the fidelities here are calculated for
\(r_1 = 10\), and \(r_2 = 9.5\), and the fields
are depicted in Fig. \ref{fig-fields}. As seen in the figure, for moderate values of the distributed noise \(r_3\),
the maximal fidelities are obtained
around \(t=3T/4\). The local noise is assumed to be absent here. The dip in the fidelity curve around \(t=3T/4\) for the very high noise (\(r_3=0.9 r_1\)) case,
is due to the fact that the energy gap between the 1st excited state and the 2nd excited state becomes comparable to
that between the 3rd and the 4th.
Again,
\(T\) is a time that satisfies Eq.
(\ref{eq-roti-kapra-aur-makaan}), which with our chosen parameters mean
\(T \gg 7 \times 10^6 \hbar/\lambda\).
There is no appreciable decrease in the fidelity upto about \(r_3=0.5 r_1\).
The horizontal line at \(2/5\) denotes the limit above which the Bell gate fidelity is quantum.
}
\label{fig-bell}
 \end{center}
\end{figure}

\subsection{Protocol for the Bell gate}

In the case of the Bell gate, the encoding is as in Eq.
(\ref{eq-dui}). Here, two qubits that are initially in the state
\begin{equation}a_{00}\ket{00}+a_{11}\ket{11} + a_{01}\ket{01}+a_{10}\ket{10}
\end{equation}
 (in the logical basis),
should evolve into the state
\begin{equation}a_{00}\ket{\phi^+} - a_{11}\ket{\phi^-} + a_{01}\ket{\psi^+} - a_{10}\ket{\psi^-}.
\end{equation}
Using the encoding in Eq. (\ref{eq-dui}), the two qubits are
initially in the state
\begin{equation}
\label{eq-haora2}
a_{00}\ket{G(0)}+a_{11}\ket{E_1(0)} + a_{01}\ket{E_2(0)}+a_{10}\ket{E_3(0)}.
\end{equation}
Again, adiabatic changes in the fields in the QNN Hamiltonian up
to a certain time \(t=T\), changes
 the state in Eq. (\ref{eq-haora2})
into the state
\begin{eqnarray}
\label{eq-champa2}
a_{00}e^{i\Phi _0}\ket{G(T)}+a_{11}e^{i\Phi _1}\ket{E_1(T)} + a_{01}e^{i\Phi _2}\ket{E_2(T)} \nonumber \\
+a_{10} e^{i\Phi _3}\ket{E_3(T)}.
\end{eqnarray}
Our strategy in this case is again to change the fields in such a way that the final (time evolved) state in Eq. (\ref{eq-champa2})
is as close as possible to the Bell rotated state
\begin{eqnarray}
    a_{00} \frac{\ket{G(0)} + \ket{E_1(0)}}{\sqrt{2}}
 -  a_{11} \frac{\ket{G(0)} - \ket{E_1(0)}}{\sqrt{2}} \nonumber \\
 +  a_{01} \frac{\ket{E_2(0)} + \ket{E_3(0)}}{\sqrt{2}}
 -  a_{10} \frac{\ket{E_2(0)} + \ket{E_3(0)}}{\sqrt{2}}.
\end{eqnarray}

\subsection{Fidelity of a gate}
\label{subsec-nikot}

The fidelity \(f\) of a gate is defined as the overlap between
the required output state $\ket{\Psi}$ of the gate and the actual final
state \(\ket{\Psi_{out}}\), averaged over the Hilbert space of input states \(\ket{\psi}\):
\begin{equation}f = \int d \left(\ket{\psi}\right) |\braket{\Psi|\Psi_{out}}|^2.
\end{equation}
Note that both the ideally required output
$\ket{\Psi}$, and the actual final state \(\ket{\Psi_{out}}\),
depends on the input state \(\ket{\psi}\).

We compare the fidelities of our gates to the ``classical''
fidelity, a term which is commonly used in the following context:
Suppose that a quantum gate takes a \(d\) level quantum system as
its input. Consider a situation where,
 instead of using the quantum gate, one uses the strategy
of measuring the input (thus making the information in the quantum input as classical), and then preparing an output from the
information obtained from the measurement on the input. The maximal fidelity that is obtainable in this way is said to be the classical
fidelity of the gate. Note that the only parameter of the quantum gate that is used here is the dimension of the input space
of the gate. The classical fidelity of a quantum gate that takes \(d\) level systems at its input is (see e.g. \cite{ref-ekhanei-ki-prothhom})
\begin{equation}\frac{2}{d+1}.
\end{equation}

\section{Fidelities of the \({\cal H}\) and Bell gates} \label{sec-fidelity}
We now study the fidelities for the \({\cal H}\) and the Bell
gates. Starting with the \({\cal H}\) gate, we investigate the behavior of the fidelity as
a function of time, for an exemplary set of values of the
parameters in the QNN Hamiltonian. We consider the influence of
distributed noise in Fig. \ref{fig-hadamard}, whereas in Fig.
\ref{fig-local_noise} we show the fidelity for different levels of
local noise. Notice that in both cases of noise, even substantial increases
in the noise level do not change the fidelity very much. Moreover,
there is a large region of the time axis where the fidelity is
larger than the classical limit \(2/3 \approx 0.667\).

We have so far considered the influences of local noise and distributed
noise separately. In a more realistic scenario, however, the two
sources of noise act simultaneously. We investigate this situation in Fig. \ref{fig-loc+glob}.
We note again that even substantially increasing both the
noise levels does not affect the fidelity in a dramatic way.

The corresponding calculations for the Bell gate lead to
qualitatively similar results. The values obtained for the
fidelities, for exactly the same system parameters as for the
\({\cal H}\) gate in Fig. \ref{fig-hadamard},
 are displayed in Fig.
\ref{fig-bell}. Note that the classical limit in this case is
\(2/5 = 0.4\).

\begin{figure}[t]
\begin{center}
\includegraphics[width=0.9\linewidth]{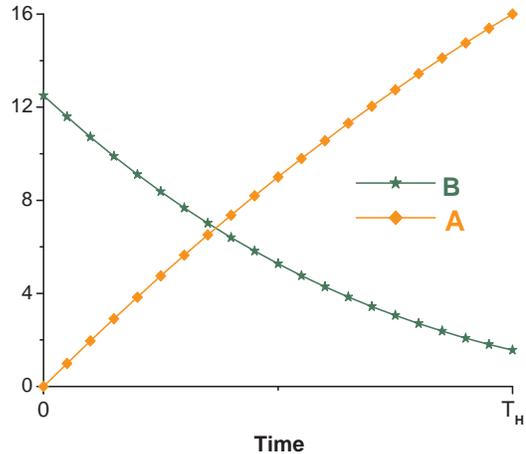}
\caption{(Color online.) The adiabatic change in the fields that effects the \({\cal H}\) and Bell gates as shown
in Figs. \ref{fig-hadamard},\ref{fig-local_noise}, \ref{fig-loc+glob}, and \ref{fig-bell}, as implemented
in a fountain trap, as well as in the implementation of the \({\cal H}\) gate in Fig. \ref{fig-hadharmonic} in a harmonic trap.
The fields are \(A(t) \lambda\) and \(B_1(t) \lambda= 10^{-5} B(t) \lambda\) and \(B_2(t) \lambda= 10^{-6} B(t) \lambda\), where
\(A(t)\) and \(B(t)\) are as shown in the figure.
For this choice of the fields, adiabaticity requires that
\(T_H \gg 7 \times 10^6 \hbar/\lambda\).
This time \(T_H\) corresponds to the time at which the fidelity of the \({\cal H}\) gate, for \(r_3 =0\), attains its maximum.
}
\label{fig-fields}
 \end{center}
\end{figure}

The changes of the fields that we make for the above implementation of the gates are the same for both the gates, and are shown in Fig.
\ref{fig-fields}.

\begin{figure}[t]
\begin{center}
\includegraphics[width=0.9\linewidth]{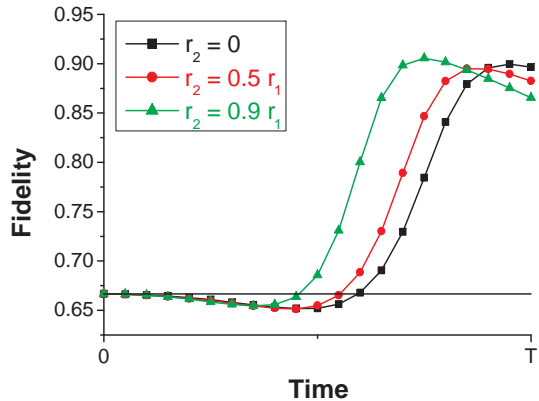}
\caption{(Color online.) Fidelity of the \({\cal H}\) gate, as a function of time, in a harmonic confinement.
The fidelities are calculated for
\(r_1 = 10\), and
the fields
are as in
Fig. \ref{fig-fields}.
\(T\) is a time that satisfies Eq.
(\ref{eq-roti-kapra-aur-makaan}), which with our chosen parameters mean
\(T \gg 7 \times 10^6 \hbar/\lambda\).
Note that the distributed noise parameter is now \(r_2\), in contrast to that in Figs. \ref{fig-hadamard},
\ref{fig-local_noise}, \ref{fig-loc+glob}, and \ref{fig-bell}. The local noise is assumed to be absent.
Again the fidelities do not change appreciably with the increase of the noise level \(r_2\).
The horizontal line at \(2/3\) denotes the limit above which the gate fidelity is quantum.
}
\label{fig-hadharmonic}
 \end{center}
\end{figure}

The gate fidelities as shown in Figs. \ref{fig-hadamard}, \ref{fig-local_noise}, \ref{fig-loc+glob}, and
\ref{fig-bell}, are for the case when \(r_1 \approx r_2 \gg r_3\),
and as shown in Ref. \cite{ager-paper, ager-paper-abar}, the
latter requirement cannot be met in a harmonic confinement of the
ions. Many experimental strategies, however, consider a harmonic
confinement, in which case one has \(r_1 \gg r_2, r_3\)
\cite{ager-paper,ager-paper-abar}, and as we show in Fig.
\ref{fig-hadharmonic}, one can implement a noise resistant \({\cal
H}\) gate in such a trap.

Let us note here that in all the above figures, where fidelities of gates are plotted with respect to time, the curves for the fidelities have
small curvatures at and around the positions of maximum fidelities. This implies that in an implementation of the
presented protocols, small errors in the time of measurement (of the fidelity), does not affect the gate fidelities appreciably.


\section{Adiabaticity and the avoided crossings}
\label{sec-adiabaticity}

The above calculations were performed by keeping in mind that we must
respect the adiabaticity condition. As we have noted before, the
adiabaticity condition demands that we should have
\begin{equation}
T \gg \hbar
\frac{\parallel \frac{d}{ds}H(s) \parallel}{g(s)^2}.
%
\end{equation}

\begin{figure}[t]
\begin{center}
\includegraphics[width=0.9\linewidth]{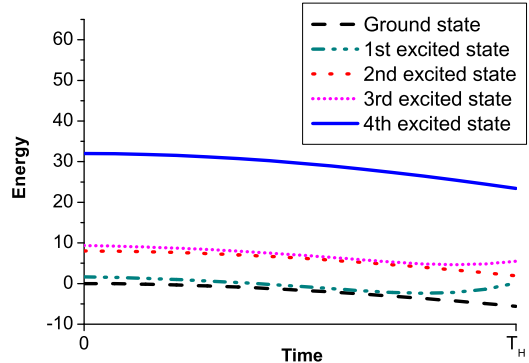}
\caption{(Color online.) Distribution of the five
lowest energy levels for the time evolution (with the system parameters being just as in Fig. \ref{fig-hadamard},
with \(r_3 =0\)), up to the point of
maximal fidelity for the \({\cal H}\) gate in Fig. \ref{fig-hadamard} (for \(r_3=0\)). The maximal fidelity for the Bell gate in Fig.
\ref{fig-bell} is obtained not long after that of the \({\cal H}\) gate in Fig. \ref{fig-hadamard}.
Note that the energy gap between the ground state and
first excited state, as well as that for the second excited and third
excited state,  are scaled up by a factor of 300 (in the figure), for better
visibility. Also the actual energy gaps as shown in the figure are to be multiplied by \(\lambda\), to have
the correct unit and value.
}
\label{fig-energy}
 \end{center}
\end{figure}

For the case of the one-qubit gate considered, there are two
energy levels involved. They are respectively the ground and the
first excited state of the whole system (the QNN). In the case of
the two-qubit gate we considered, there are four energy levels
involved. They are the ground state, and the first, second and
third excited states of the whole system. The maximal gate
fidelities are reached after the system passes through a
``double'' avoided crossing. One of the avoided crossings is
between the ground state and the first excited state, while the
other is between the second and the third excited states. They
appear almost at the same time. In Fig. \ref{fig-energy}, we show
the dynamics of the five lowest energy eigenvalues, when
\(r_1=10\), \(r_2 = 9.5\), and \(r_3=0\), and the fields as in
Fig. \ref{fig-fields}. A typical value for the energy gap at the
avoided crossing is \(\lambda \times \mbox{0.03}\).  Note that for
the adiabatic transfer in the implementation of the \({\cal H}\) gate, the three lowest levels are the relevant ones,
while for the Bell gate implementation,
the five lowest levels are relevant. For
the above values of \(r_1\) and \(r_2\), and for values of \(r_3\)
up to \(\approx 0.9r_1\), the typical energy gap (at the
avoided crossing), remains approximately at \(\lambda \times \mbox{0.03}\). For higher values of the distributed
noise level \(r_3\),
i.e. for the case when \(r_1 \approx r_2 \approx r_3\), this gap
collapses, and hence it is no more possible to implement the gates
in the presented way.

We denote by \(T_H\), the point of time at which the maximal fidelity is
reached for the \({\cal H}\) gate, for vanishing distributed noise \(r_3\), and vanishing local noise
\(\varepsilon\), in the fountain trap. In the same situation, the maximal fidelity of the Bell gate is
attained approximately at the
same point of time. The avoided crossing is approximately at
\(3T_H /4\). Adiabaticity demands that
\begin{equation}T_H \gg 7 \times 10^6 \frac{\hbar}{\lambda}.
\end{equation}

\section{Discussion} \label{sec-summary}

We suggest a realization of universal quantum computing on an
experimentally viable system of \emph{distributed} qubits: The
qubits are encoded in the (low) energy levels of the whole system.
As in classical neural networks, where the distributed storage of
classical information allows for robustness to noise, we show that
our quantum system is resistant to high levels of noise. The one-
and two-qubit quantum gates described in this paper are realized
via adiabatic passage of the system from one set of energy
eigenstates to another set of corresponding eigenstates. The
adiabatic transfer is effected by a slow change of parallel and
transverse fields. We perform numerical simulations to obtain the
gate fidelities, and show that for a certain slow change of the
fields, the gate fidelities are indeed much higher than their
classical limits. We also observe that the fidelities typically
have small curvatures near their maxima. Therefore, the gate
fidelities will not change appreciably for small errors, in the
time of measuring of the fidelities, in the experiments. The
scalability issue is like in other proposals and experiments in
ion-trap quantum computing \cite{seidelin}, and may potentially be
overcome by connecting mesoscopic clusters of trapped ions by
''flying'' qubits.

In this paper, we have considered the implementations of two gates:
A one-qubit gate, which we have called the \({\cal H}\) gate,
because of its similarity to the Hadamard gate, and a two-qubit
gate, which we call the Bell gate, because the output states for
an input computational basis, are the Bell states (up to phases).
For the implementation of the \({\cal H}\) gate, there are two
energy levels involved: the ground state and the first excited
state of the whole system (the quantum neural network). For the
implementation of the Bell gate, there are four energy levels
involved: the ground state, and the first three excited states of
the whole system. We observe that the maximal gate fidelities are
reached after the system passes through a ``double'' avoided
crossing. One of the avoided crossings is between the ground state
and the first excited state, while the other is between the second
and third excited states, and they appear almost at the same point
of time. We find the condition under which the adiabaticity is
realized.

The approach to noise management in both classical and quantum computational networks can broadly be divided into
two categories: ``active'' and ``passive'' schemes. In a typical \emph{active} noise-management scheme, the
classical or quantum information is encoded in several bits or qubits, so that if noise occurs, it can be detected, and subsequently
corrected.
The active scheme is followed, for example, for data storage in a typical compact disc
of our classical computers. There is also a corresponding theory of quantum error correction and
fault tolerant quantum computation \cite{faulttol}, that acts in an active way to correct errors. Typically, such a scheme
will require encoding the quantum information into multiqubit entangled states, detecting any possible noise effect by multiqubit
measurements, and reverting the effect of noise by a multiqubit unitary (which can be replaced by single-qubit and two-qubit unitaries).

A typical \emph{passive} noise-management tries to identify a
system that is already resistant to errors. Classical systems that
manage noise effects in this way is are neural network models of
brains. Examples of such noise-management in quantum circuits
include the theories of decoherence-free subspaces
\cite{decfreesubsp} and topological quantum computation
\cite{topological}. Our model of quantum computation by using a
quantum neural network also falls into this category. A similarity
between our scheme and a typical active noise-management scheme
\cite{faulttol} is that both require to deal with several qubits:
While the methods in Refs. \cite{faulttol} encode the quantum
information in several qubits so that a possible noise effect is
detectable and reversible, our quantum neural network model of
quantum computation encodes the quantum information in a
mesoscopic number of qubits so that the system is resilient
towards noise. However, our model does not require encoding into
multiqubit entangled states, or measurements onto multiqubit
bases. Moreover, the step to revert the effect of noise is absent.

\begin{acknowledgments}

We acknowledge helpful comments from J{\"u}rgen Eschner, and support from the DFG (SFB 407, SPP 1078, SPP 1116),
ESF QUDEDIS, Spanish MEC (FIS-2005-04627, Consolider Project QOIT,
and Ramon y Cajal), EU IP SCALA, DAAD (German Academic Exchange
Service), and the Ministry of Education of the Generalitat de
Catalunya.

\end{acknowledgments}


\begin{thebibliography}{99}

\bibitem{shor} P. Shor,
SIAM J. Sci. Statist. Comput.
\textbf{26},
1484
(1997) (quant-ph/9508027).

\bibitem{grover} L. Grover, in
\emph{Proceedings of the 28th Annual ACM Symposium on the
Theory of Computing (STOC)}, pp. 212–-219 (1996) (quant-ph/9605043);
Phys. Rev. Lett. \textbf{79}, 325 (1997).

\bibitem{faulttol} P. Shor,  in \emph{37th Symposium on Foundations of Computing}, pp. 56--65 (IEEE Computer Society Press, 1996)
(quant-ph/9605011); E. Knill and R. Laflamme,
quant-ph/9608012; E. Knill, R. Laflamme and W. Zurek,
quant-ph/9610011; Proc. R. Soc. Lond. A \textbf{454}, 365 (1998) (quant-ph/9702058);
D. Aharonov and M. Ben-Or, quant-ph/9611025; quant-ph/9906129; A.Y. Kitaev, in \emph{Proc. 3rd Int. Conf. of Quantum Communication and
Measurement} (Plenum, NY, 1997), and references therein.

\bibitem{decfreesubsp} L.M. Duan and G.C. Guo, Phys. Rev. Lett. \textbf{79}, 1953 (1997); P.
Zanardi and M. Rasetti, Phys. Rev. Lett. \textbf{79}, 3306 (1997); D.A.
Lidar, I.L. Chuang, and K.B. Whaley, Phys. Rev. Lett. \textbf{81}, 2594
(1998).

\bibitem{noiselessubsys} E. Knill, R. Laflamme, and L. Viola, Phys. Rev. Lett. \textbf{84}, 2525
(2000).

\bibitem{dyndecoupling} L. Viola and S. Lloyd, Phys. Rev. A. \textbf{58}, 2733 (1998); L. Viola,
E. Knill, and S. Lloyd, Phys. Rev. Lett. \textbf{82}, 2417 (1999); P. Zanardi, Phys. Lett. A \textbf{258}, 77 (1999); D. Vitali and P.
Tombesi, Phys. Rev. A \textbf{65}, 012305 (2002).

\bibitem{topological} A.Y. Kitaev,  Annals Phys. \textbf{303}, 2 (2003) (quant-ph/9707021); M.H. Freedman,  A. Kitaev, and Z. Wang,
 Commun. Math. Phys. \textbf{227}, 587 (2002) (quant-ph/0001071);
 M.H. Freedman, M. Larsen, and Z. Wang, quant-ph/0001108; M.H. Freedman, quant- ph/0003128;
 M.H. Freedman, A. Kitaev, M.J. Larsen, and Z. Wang, quant-ph/0101025;
C. Mochon, Phys. Rev. A \textbf{67}, 022315 (2003); B. Dou{\c
c}ot, M.V. Feigel'man, L.B. Joffe, and A.S. Ioselevich,  Phys.
Rev. B \textbf{71}, 024505 (2005),
and references therein.

\bibitem{gyamiti} P. Zanardi and M. Rasetti, Phys. Lett. A \textbf{264}, 94 (1999); J. Pachos, P. Zanardi, and
M. Rasetti, Phys. Rev. A \textbf{61}, 10305 (2000);
J.A. Jones, V. Vedral, A. Ekert, and G. Castagnoli, Nature \textbf{403}, 869 (2000); A. Ekert, M. Ericsson, P. Hayden, H. Inamori,
J.A. Jones, D.K.L. Oi, and V. Vedral, J. Mod. Opt. \textbf{47}, 2501 (2000), and references therein.

\bibitem{neuralnets} D.J. Amit, \emph{Modeling Brain Functions: The World of Attractor Neural Networks} (Cambridge
University Press, Cambridge, 1989); M. M{\' e}zard, G. Parisi, and M.A. Virasoro,
\emph{Spin Glass Theory and Beyond: An Introduction
to the Replica Method and Its Applications} (World Scientific, Singapore, 1987).

\bibitem{eita-Hamming} T.M. Cover and J.A. Thomas, \emph{Elements of Information Theory} (Wiley, New York, 1991).

\bibitem{eita-Wineland_at_Boulder} D. Leibfried, E. Knill, S. Seidelin, J. Britton, R.B. Blakestad,
J. Chiaverini, D.B. Hume, W.M. Itano, J.D. Jost, C. Langer, R. Ozeri, R. Reichle, and D.J. Wineland,
Nature \textbf{438}, 639
(2005).

\bibitem{eita-Blatt} H. H{\" a}ffner, W. H{\" a}nsel,
C.F. Roos, J. Benhelm, D. Chek-al-kar, M. Chwalla, T. K{\" o}rber, U.D. Rapol, M. Riebe, P.O. Schmidt, C. Becher, O. G{\" u}hne,
W. D{\" u}r, and R. Blatt,
Nature \textbf{438}, 643
(2005).



\bibitem{Porras-Cirac} D. Porras and J.I. Cirac, Phys. Rev. Lett. \textbf{92}, 207901
(2004).

\bibitem{Wunderlich} C. Wunderlich, in \emph{Laser Physics at the Limit} (Springer, Heidelberg, 2002), p. 261
(quant-ph/0111158).


\bibitem{review} M. Lewenstein, A. Sanpera, V. Ahufinger, B. Damski, A Sen(De),
and U. Sen, Adv. in Phys. \textbf{56}, 243
 (2007) (cond-mat/0606771).


\bibitem{ager-paper-abar} M. Pons, V. Ahufinger, C. Wunderlich, A. Sanpera,
S. Braungardt, A. Sen(De), U. Sen, and M. Lewenstein, Phys. Rev.
Lett. \textbf{98}, 023003 (2007).


\bibitem{born}  M. Born and V. Fock,
Zeitschrift
f\"ur Physik
\textbf{51},
165
(1928).

\bibitem{messiah} A. Messiah, \emph{Quantum Mechanics}, (John Wiley \& Sons, New York, 1958).


\bibitem{Pancharatnam-birat-byapar} S. Pancharatnam, Proc. Indian Acad. Sci. \textbf{44}, 247 (1956).

\bibitem{Berry} M.V. Berry, Proc. R. Soc. Lond. A \textbf{392}, 84 (1984).

\bibitem{Wilczek} A. Shapere and F. Wilczek (Eds.),
\emph{Geometric Phases in Physics} (World Scientific, Singapore, 1998).





\bibitem{farhi} E. Farhi and S. Gutmann, Phys. Rev. A \textbf{57},
2403 (1998).


\bibitem{aro_onekei_achhey} E. Farhi, J. Goldstone, S. Gutmann, and M. Sipser, quant-ph/0001106;  W. van Dam, M. Mosca, and U. Vazirani,
in \emph{Proceedings of the 42nd Annual Symposium on Foundations of Computer Science}, pp. 279--287 (2001) (quant-ph/0206003).

\bibitem{ina-mina-dika} D. Aharonov,  W. van Dam, J. Kempe, Z. Landau, S. Lloyd, and O. Regev, quant-ph/0405098.










\bibitem{ager-paper}M. Pons, V. Ahufinger, C. Wunderlich, A. Sanpera,
 and M. Lewenstein,  cond-mat/0512606.

\bibitem{W-state} W. Dur, G. Vidal, and J. I. Cirac, Phys. Rev. A \textbf{62}, 062314 (2000).



\bibitem{ref-ekhanei-ki-prothhom} M. Horodecki, P. Horodecki, and R. Horodecki,
Phys. Rev. A \textbf{60}, 1888
(1999).

\bibitem{seidelin} See e.g., S. Seidelin, J. Chiaverini, R. Reichle, J. J. Bollinger, D. Leibfried, J. Britton, J. H. Wesenberg, R. B. Blakestad, R. J. Epstein, D. B. Hume, W. M. Itano, J. D. Jost, C. Langer, R. Ozeri, N. Shiga, and D. J. Wineland, Phys. Rev. Lett. \textbf{96}, 253003
(2006).

\bibitem{phuler-upottyoka} D. Deutsch, A. Barenco, and A. Ekert, Proc. R. Soc. Lond. A {\bf 449}, 669 (1995), and
references therein.

\end{thebibliography}
\end{document}